# Machine-learning based three-qubit gate for realization of a Toffoli gate with cQED-based transmon systems

Sahar Daraeizadeh[1, 2], Shavindra P. Premaratne [2], Xiaoyu Song[1], Marek Perkowski[1], Anne Y. Matsuura[2]

[1] Department of Electrical and Computer Engineering, Portland State University, Portland, Oregon 97201, USA

[2] Intel Labs, Intel Corporation, Hillsboro, Oregon 97124, USA

We use machine learning techniques to design a 50 ns three-qubit flux-tunable controlled-controlled-phase gate with fidelity of >99.99% for nearest-neighbor coupled transmons in circuit quantum electrodynamics architectures. We explain our gate design procedure where we enforce realistic constraints, and analyze the new gate's robustness under decoherence, distortion, and random noise. Our controlled-controlled phase gate in combination with two single-qubit gates realizes a Toffoli gate which is widely used in quantum circuits, logic synthesis, quantum error correction, and quantum games.

## I. Introduction

Circuit quantum electrodynamics (cQED) systems [1-3] utilizing transmons [4-5] are one of the potential candidates for realizing quantum computers [6], with qubit coherence times of hundreds of microseconds [7] and the potential to scale up facilitated by quantum error correction schemes [8-9] . Here, we theoretically design a three-qubit controlled-controlled phase (CCPhase) gate with fidelity of >99.99% for nearest-neighbor transmons with resonator couplings [10].

Multiple-qubit-controlled-phase gates in transmons are typically designed by detuning the qubit transition frequencies to approach the avoided-level-crossing regions. In this regime, state mixing or level shifting due to non-computational quantum levels allows non-uniform phase collection within the computational subspace. This gives rise to entangling operations between qubits [11-16]. Finding the optimal frequency detuning for transmons to achieve the desired avoided level crossings between the system energy levels is a complex task which can benefit from machine learning (ML) approaches [17-19]. Designing quantum gates and optimized control pulses using ML techniques and optimization theory has been demonstrated for various quantum systems [20-23]. We model the quantum gate design problem as a supervised ML exercise, by adjusting the system control parameters to converge on the target gate unitary [18]. In this model, the system parameters can be learned from the training set which is the desired unitary matrix, and the cost function is the gate fidelity.

## II. Toffoli gate realization

The Toffoli gate has broad applications in many quantum circuits. The best-known decomposition of the Toffoli (controlled-controlled-NOT) gate using standard single- and two-qubit gates [24] requires multiple single-qubit gates ($H$, $T$, and $T^\dagger$) and 6 CNOT gates as shown in Fig. 1(a). In this decomposition, at least two of the CNOT gates are applied to non-neighbor qubits which results in addition of four SWAP gates in a nearest-neighbor architecture. There is another decomposition of the Toffoli gate based on five two-qubit gates [25] as depicted in Fig. 1(b) where non-standard two-qubit gates such as controlled-V and controlled-$V^\dagger$ gates are required where $V^2 = X$, and $VV^\dagger = I$. In other words, controlled-V and controlled-$V^\dagger$ gates can be represented by $c\sqrt{\text{NOT}}$, and $c\sqrt{\text{NOT}}^\dagger$, respectively. The $c\sqrt{\text{NOT}}$, and $c\sqrt{\text{NOT}}^\dagger$ gates can be realized using controlled-rotation flux-tunable gates in transmons in cQED systems, however, the circuit shown in Fig. 1b, requires two extra SWAP gates to perform a controlled-rotation gate between non-neighbor qubits in a nearest-neighbor architecture.

The decomposition of the Toffoli gate based on single- and two-qubit gates is costly. Another proposed decomposition of the Toffoli gate is based on a three-qubit CCPhase gate and two single-qubit gates (Hadamard or single-qubit rotation gates) as shown in Fig. 1(c). Here, we show that a Toffoli gate can be realized in only 90 ns for a resonator-coupled nearest-neighbor transmon system utilizing the single-qubit gates (20 ns) [10], and our high fidelity CCPhase gate (50 ns) with realistic frequency detuning sequences and system parameters.

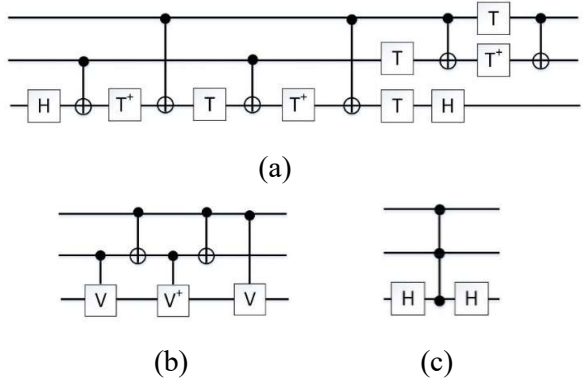

**FIG. 1.** Toffoli gate decomposition circuits. (a) Toffoli circuit based on standard single-qubit and two-qubit gates. (b) Toffoli gate circuit based on two-qubit gates. (c) Toffoli circuit based on single-qubits gates and three-qubit controlled-controlled-phase gate

The CCPhase gate is designed to collect a $\pi$ phase only on the $|111\rangle$ computational state (i.e. when all three qubits are in $|1\rangle$ state). For the CCPhase gate simulation, we consider the lowest four energy levels (labeled $|0\rangle$ to $|3\rangle$) to ensure system evolution within the full three-excitation manifold [15]. However, the cost function evaluation for the ML approach is performed only within the qubit subspace. The unitary operation of the ideal CCPhase gate in matrix form is:

$$U_{\text{ideal}} = \begin{bmatrix} 1 & 0 & 0 & 0 & 0 & 0 & 0 & 0 \\ 0 & 1 & 0 & 0 & 0 & 0 & 0 & 0 \\ 0 & 0 & 1 & 0 & 0 & 0 & 0 & 0 \\ 0 & 0 & 0 & 1 & 0 & 0 & 0 & 0 \\ 0 & 0 & 0 & 0 & 1 & 0 & 0 & 0 \\ 0 & 0 & 0 & 0 & 0 & 1 & 0 & 0 \\ 0 & 0 & 0 & 0 & 0 & 0 & 1 & 0 \\ 0 & 0 & 0 & 0 & 0 & 0 & 0 & -1 \end{bmatrix} \quad (1)$$

where the ordering of the states is $|000\rangle$ to $|111\rangle$ in binary increments.

### III. Simulation of the system dynamics

The effective Hamiltonian for our model with $n$ transmons, when the coupling resonators are not populated, can be described as follows [26]:

$$\mathcal{H} = \sum_{k=1}^{n-1} \widetilde{\mathcal{H}}_c^{(k,k+1)} + \sum_{k=1}^{n} \widetilde{\mathcal{H}}_t^{(k)} \quad (2)$$

Here, the Hamiltonian of each transmon $k$ is:

$$\widetilde{\mathcal{H}}_t^{(k)} \equiv \sum_j \widetilde{\omega}_j^{(k)} |j\rangle_{(k)}\langle j| \quad (3)$$

where $\widetilde{\omega}_{(j)}^{(k)}$ is the dressed transition frequency associated with the $k^{\text{th}}$ transmon at energy level $j$ and is given by

$$\widetilde{\omega}_j^{(k)} \equiv j\omega_q^{(k)} + \frac{\delta_k}{2}(j-1)j + \frac{jg_k^2}{\omega_q^{(k)} - \omega_r + (j-1)\delta_k} \quad (4)$$

where $\omega_q^{(k)}$ is the bare transition frequency associated with qubit $k$, $g_k$ is the coupling strength between transmon $k$ and the connected resonator, and $\omega_r$ represents the frequency of the coupled resonator. The last term in Eq. 4 is repeated for each transmon with appropriate modifications depending on the number of coupled resonators.

For any pair of coupled transmons via a resonator, we estimate the direct coupling between two transmons $(k, k+1)$ as:

$$\widetilde{\mathcal{H}}_c^{(k,k+1)} = \sum_{j_k, j_{k+1}} \sqrt{j_k + 1}\sqrt{j_{k+1} + 1} J_{j_k, j_{k+1}}(|j_k, j_{k+1} + 1\rangle\langle j_k + 1, j_{k+1}| + |j_k + 1, j_{k+1}\rangle\langle j_k, j_{k+1} + 1|) \quad (5)$$

where $J_{j_k, j_{k+1}}$ is the direct coupling between level $j_k$ from the $k^{\text{th}}$ transmon and level $j_{k+1}$ from the $(k+1)^{\text{th}}$ transmon.

$$J_{j_k, j_{k+1}} = \frac{g_k g_{k+1}\left(\omega_q^{(k)} + \delta_k j_k - \omega_r + \omega_q^{(k+1)} + \delta_{k+1} j_{k+1} - \omega_r\right)}{2\left(\omega_q^{(k)} + \delta_k j_k - \omega_r\right)\left(\omega_q^{(k+1)} + \delta_{k+1} j_{k+1} - \omega_r\right)} \quad (6)$$

where $\delta_k$ and $\delta_{k+1}$ are the anharmonicity values associated with transmons $k$ and $k+1$, respectively.

Using the time-dependent Hamiltonian of the system, the time evolution equation of the system is solved to achieve the unitary transformation $U$:

$$U(t) = \exp\left\{-\frac{i}{\hbar}\int_0^t \mathcal{H}(\tau)\, d\tau\right\} \quad (7)$$

Here $t$ is the time, $\mathcal{H}$ is the Hamiltonian of the system, and $\hbar$ is the reduced Planck's constant. To solve Eq. 7, we employ Trotterization [27]. Hence, the final unitary transformation is estimated as follows [28]:

$$U(t_k) = U_k U_{k-1} U_{k-2} \ldots U_2 U_1 U_0 \quad (8)$$

Here $U_i$ for $i = \{0, 1, \ldots, k\}$ is calculated using Eq. 7 for the newly time-independent Hamiltonian at each timestep $i$, where $U_0 = I$, and $k$ is the total number of steps. The Trotter step size is $T/k$, where $T$ is the gate evolution time. In our simulations, the Trotter step size was 100 ps.

When solving the time evolution equation, we considered a smaller subspace to reduce the computational expenses. The Hamiltonian for $n$ transmons with four energy levels spans a $4^n$-dimensional Hilbert space. For a system composed of three transmons ($n=3$) the Hamiltonian is a $64\times 64$ matrix operator. Solving the Schrödinger equation for this large operator is computationally expensive, and there are numerous energy levels that have a minimal impact on the evolution for the gate of interest. Thus, we project this larger Hamiltonian to a smaller subspace where at most three excitations are allowed, resulting in a $20\times 20$ matrix [18]. The 20 states considered are $\{|000\rangle, |001\rangle, |002\rangle, |003\rangle, |010\rangle, |011\rangle, |012\rangle, |020\rangle, |021\rangle, |030\rangle, |100\rangle, |101\rangle, |102\rangle, |110\rangle, |111\rangle, |120\rangle, |200\rangle, |201\rangle, |210\rangle, |300\rangle\}$.

The reduced Hamiltonian is evolved based on the qubit transition frequencies. The resulting unitary is projected [18] to the $8\times 8$ computational subspace that includes the states $\{|000\rangle, |001\rangle, |010\rangle, |011\rangle, |100\rangle, |101\rangle, |110\rangle, |111\rangle\}$. Single-qubit phase compensation [14], [17-18] is performed on this resultant unitary using the diagonal compensation matrix

$$M = e^{-i\theta_0}\,\mathrm{diag}(1, e^{-i\theta_1}, e^{-i\theta_2}, e^{-i(\theta_1+\theta_2)}, e^{-i\theta_4}, e^{-i(\theta_1+\theta_4)}, e^{-i(\theta_2+\theta_4)}, e^{-i(\theta_1+\theta_2+\theta_4)}) \quad (9)$$

where $\theta_0$ represents the global phase, and $\theta_1$, $\theta_2$, and $\theta_4$ represent the relative single qubit phases of states $|001\rangle$, $|010\rangle$, and $|100\rangle$, respectively.

The single qubit phases are cancelled out by multiplying matrix $M$ with the projected unitary in the computational subspace:

$$U_{\text{final}} = U_{\text{proj}} \times M \quad (10)$$

Finally, we calculate the gate fidelity considering unitarity $U_{\text{final}}$ and its closeness to the target ideal operation from Eq. 1 as follows [29]:

$$\mathcal{F} = \frac{\mathrm{Tr}(U^\dagger U) + |\mathrm{Tr}(U_{\text{ideal}}^\dagger U)|^2}{d(d+1)} \quad (11)$$

where $d = 2^3$ is the dimensionality of the computational subspace.

## IV. CCPhase gate design using machine learning methods

There are many machine learning and optimization algorithms one can choose to solve the optimal control problem. We design the system parameters to realize the CCPhase gate by combining two learning methods: (1) A machine learning method based on differential evolution [30] named Subspace-Selective Self-Adaptive Differential Evolution (SUSSADE) [17-18], (2) our new local search algorithm. In both learning procedures, the gate fidelity as shown in Eq. 11 is considered as the fitness function to achieve the optimal control parameters for the given ideal unitary. During the learning procedure all parameters are assumed to be fixed, except the frequency detuning sequences of transmons.

In our simulations, the resonator-transmon couplings are set to $g = 0.2$ GHz, and anharmonicity of each transmon was $\delta = -0.3$ GHz. The three transmons (labeled Left, Middle, Right) with reference transition frequencies set to 5, 6, and 7 GHz, realize an identity operation with fidelity 99.9%. Transmons L and M are coupled with a 8.05 GHz resonator, and transmons M and R are coupled with a 8.2 GHz resonator.

To reduce the search space during the learning procedure, the reference transition frequencies of the qubits are set closer during the ML algorithms search; $f_L = 5.61$ GHz, $f_M = 6$ GHz, and $f_R = 6.39$ GHz, repectively. The maximum frequency detuning ranges permitted from the reference frequency of each qubit are set to [0, 0.5), (-0.5, 0.5), and (-0.5, 0], for qubits L, M, and R, respectively. These constraints help further reduce the search space and increase efficiency of the learning process by removing trial of detuning values far away from the interaction region.

Note, we further impose the following constraints during learning to ensure that the optimal frequency detuning sequences are experimentally realistic and achievable, and that the target gate is robust. We enforce these constraints by:

1. Limiting the maximum point-to-point variation of the frequency detuning sequence of each qubit to 220 MHz to prevent undesired excitations in the quantum system. To take into account the limitations of physical signal instrumentation [31], the initial and the final points are limited to maximum point-to-point variation of 500 MHz from the initial reference transition frequencies of 5, 6, and 7 GHz.

2. Limiting the minimum difference between transition frequencies of two adjacent qubits to 0.21 GHz; primarily to prevent interactions within the single-excitation manifold.

Here we briefly describe how the SUSSADE algorithm [17-18] was used to generate the qubit transition detuning sequences. First a random population of 200 random frequency detuning sequences (chromosomes) is generated in which each sequence contains 150 frequencies (50 per qubit). For a gate timing of $T = 50$ ns, the detuning sequence of each qubit is discretized to 50 amplitudes.

After generating the initial population, we perform SUSSADE by randomly modifying the values of detuning sequences using the differential evolution operations such as mutation, crossover, and selection [18], [30]. Finally, the fidelity of the resulting final unitary is calculated using Eq. 11. For any modified detuning sequence, if the new fidelity value is larger than the initial one, the new detuning sequence survives to the next generation. This procedure repeats until we reach our choice of fidelity threshold value (99.99%) or maximum number of iterations (one million cycles). We use the Message Passing Interface to distribute the simulation to 200 nodes on a computer cluster [32] such that each node is performing a full cycle of solving the time evolution and fidelity calculation for each member of the population.

SUSSADE was successfully used to obtain the frequency detuning sequences for a 50 ns CCPhase gate with fidelity of 98.8%, but any further progress was slow. Thus, a local search algorithm was implemented to refine the detuning sequences and achieve a gate fidelity of 99.99%. Note that the local search algorithm is efficient once the search space has been reduced by other learning algorithms.

The local search algorithm consists of the following steps:

1- In the beginning of the learning process, we define the largest (100 MHz) and the smallest (1 KHz) change in frequency detuning allowed per data point. This is referred to as the optimization step size $\epsilon$. We also set the maximum number of iterations (1000), the desired fidelity (99.99%), and all constraints enforced during SUSSADE.

2- While the constraints are met and the desired fidelity or maximum number of iterations have not been reached, the following procedure is repeated:

a) A local search window is moved from the first data point toward the last data point.

b) At each window, we recursively vary the frequency detuning value up or down by the optimization step size $\epsilon$ as long as it keeps improving the gate fidelity.

c) Once the local search window has covered all data points of the detuning sequence of all qubits, we reduce $\epsilon$ for a finer grain optimization ($\epsilon_{new} = 0.1\epsilon_{old}$).

d) If the optimization is already completed for the smallest predefined $\epsilon$ during the iteration, we increase the iteration number by one, reset $\epsilon$ to the largest predefined value, and repeat from step a.

The CCPhase gate duration is set to 50 ns for evaluation, and the learning algorithms operate on 1 ns step size. However, time evolution is in Trotter steps of size 100 ps (k=500 in Eq. 8). The learned frequency detuning sequences are kept constant during each 1 ns step to obtain piecewise-constant pulse forms as shown in Fig. 2.

## V. Gate verification and impact of decoherence

Simulated quantum process tomography (QPT) was used to independently evaluate gate performance by using master equation simulations. QPT is an excellent tool to evaluate the dynamics of a quantum system due to any process [33], in this case the CCPhase gate. Given that this is QPT within simulation, state preparation and measurement errors do not affect the methodology. Hence the results from QPT enable us to fully characterize the introduced gate.

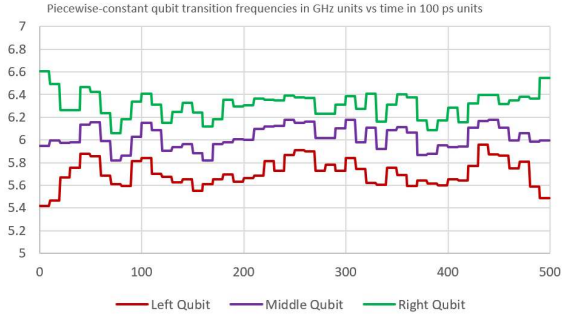

**FIG. 2.** (Color online) The learned transition frequency detuning sequences of the right, middle, and left qubits are shown from top to bottom respectively. The piecewise constant forms are generated from the learned frequency detuning sequences (50 learned data points per each transmon).

Initial verification was performed assuming no decoherence in the system by using the Von Neumann equation for time evolution:

$$i\hbar \frac{\partial \rho}{\partial t} = \mathcal{H}\rho - \rho\mathcal{H}, \quad (12)$$

where the Hamiltonian $\mathcal{H}$ is the same as that given in Eq. 2 with the number of levels in each transmon set to $j_{max} = 4$, and $\rho$ is the density matrix for the three transmon system.

The three transmon system was evolved using the generated resonance frequency detuning sequences from learning algorithms. The evolution was performed on all the initial states given by $\{I, R_x^{0.5\pi}, R_y^{0.5\pi}, R_x^{\pi}\}^{\otimes 3}|000\rangle$ resulting in 64 density matrices. Unlike experimental QPT, it was not necessary to perform quantum state tomography to reconstruct these density matrices for the final states. These results were used to perform QPT by imposing constraints that the process matrix $\chi$ must satisfy [34-35]. The $\chi$ matrix completely characterizes the underlying process and is positive-Hermitian by definition [33].

We use the following metrics as defined in Ref. [34] to evaluate the performance of the CCPhase gate:

Process fidelity: $\mathcal{F}_p = \text{Tr}(\chi^{(\text{ideal})}\chi)$ (13)

Average gate fidelity: $\mathcal{F}_g = \frac{d\mathcal{F}_p + 1}{d+1}$ (14)

Average purity: $\overline{\text{Tr}(\rho^2)} = \frac{d\,\text{Tr}(\chi^2)+1}{d+1}$ (15)

where $\chi$ is the experimentally determined process matrix, $\chi^{(\text{ideal})}$ is the ideal process matrix for the CCPhase gate, and $d = 2^3$ is the dimensionality of the computational subspace of the system. The results from evaluation are given in TABLE I.

**TABLE I.** Table of QPT metrices for simulations under different conditions

| Conditions | $\mathcal{F}_p$ | $\mathcal{F}_g$ | $\overline{\text{Tr}(\rho^2)}$ |
|---|---|---|---|
| $k_{max} = 4, T_1 = T_2 = \infty$ | 0.999 | 0.999 | 0.999 |
| $k_{max} = 3, T_1 = T_2 = \infty$ | 0.998 | 0.998 | 0.999 |
| $k_{max} = 4, T_1 = T_2 = 20\ \mu s$ | 0.995 | 0.995 | 0.991 |
| $k_{max} = 3, T_1 = T_2 = 20\ \mu s$ | 0.993 | 0.994 | 0.991 |

The simulations incorporating decoherence were performed using the Lindblad-Kossakowski form of the master equation [36-37]. The appropriate operators for the dephasing portion of the master equation were obtained as in Refs. [38-39]. For convenience in simulation, $T_1$ and $T_2$ were both set to 20 μs, assuming coherence times independent of the flux-tuning of the transmons [40]. Please refer to supplementary material for the full process matrices resulting from QPT. Comparison of results for $k_{max} = 3$ ($\{|0\rangle, |1\rangle, |2\rangle\}$ levels) and $k_{max} = 4$ ($\{|0\rangle, |1\rangle, |2\rangle, |3\rangle\}$ levels) from Table I indicates that the fourth level ($|3\rangle$) also plays a limited role in the system evolution.

## VI. Robustness evaluation

The frequency detuning sequences derived from the learning algorithms have a piecewise-constant form. To investigate the effect of first-order distortion due to control electronics, we use the following pulse reshaping method [14], [18] to smooth the frequency detuning sequences:

$$\omega_k(t) = \frac{\omega_{k_i} + \omega_{k_{i+1}}}{2} + \frac{\omega_{k_{i+1}} - \omega_{k_i}}{2}\left[\text{Erf}\left(\frac{t-\left(\frac{t_{\text{ramp}}}{2}\right)}{\sqrt{2}\sigma}\right)\right], \quad (16)$$

where $\omega_k(t)$ represents the distorted frequency detuning of qubit $k$ during $t_i \leq t \leq t_{i+1}$, and $t_i$ represents the $i^{\text{th}}$ time step. Here $\text{Erf}(t) \equiv \frac{2}{\sqrt{\pi}}\int_0^t e^{-x^2}dx$ is the error function value of $t$, $t_{\text{ramp}} = 1$ ns, and $\sigma = \frac{t_{\text{ramp}}}{4\sqrt{2}}$ [14]. The distorted sequences are shown in Fig. 3 (a), and we observed a fidelity reduction of 1.21%, which resulted in average fidelity of 98.79%.

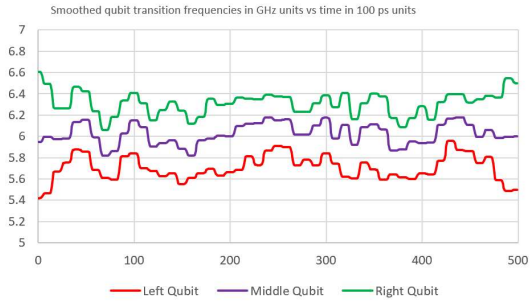

(a)

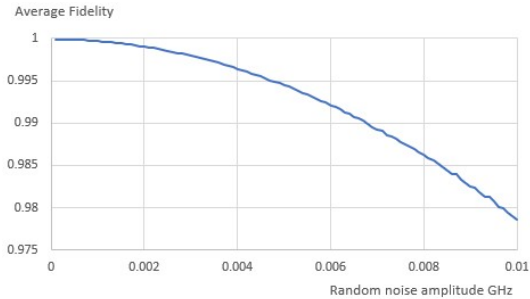

(b)

**FIG. 3.** (Color online) The effect of distortion and random noise on the learned transition frequency detuning sequences. (a) The smoothed transition frequency detuning sequences of the right, middle, and left qubits are shown from top to bottom respectively. (b) The CCPhase gate average fidelity over 10000 samples under the effect of random noise with amplitudes ranging in 0 to 10 MHz.

To investigate the effect of random noise on the CCPhase gate, we plot the average fidelity while increasing the random noise with amplitudes varying from 0 to 10 MHz. For each amplitude value, random noise is generated from a uniform distribution $(-1, 1)$, multiplied by the noise amplitude and added to the optimized detuning sequence. The latter step is repeated 10000 times and at each iteration the system Hamiltonian is evolved, and the gate fidelity is calculated. The averaged fidelity of the 10000 results is reported as the average fidelity at each noise amplitude. Fig. 3 (b) illustrates the gate robustness against random noise and demonstrates fidelity >99% with random noise amplitudes of up to 6.7 MHz.

## VII. Conclusion

We designed a robust CCPhase entangling gate with fidelity >99.99% and duration of 50 ns for quantum systems of nearest neighbor transmons coupled via resonators. We describe the gate design procedure using simulation and machine learning techniques and present a new local search algorithm for optimal quantum control applicable to small search spaces. The operation of the CCPhase gate is confirmed by a simulator in C++ that solves the Schrödinger equation for the time-dependent Hamiltonian of the system. Moreover, the gate operation is verified independently via quantum process tomography in both the presence and absence of decoherence. Gate robustness is examined using random noise injection and frequency detuning distortion. The presented theoretical gate design procedure, verification, and robustness investigation can be applied to design new gates for other quantum systems as well.

**Acknowledgements**

This work was made possible in part thanks to Portland Institute for Computational Science and its resources acquired using NSF Grant # DMS 1624776 and ARO Grant #W911NF-16-1-0307.

**Author contributions**

S. D. conceived the project, implemented a quantum simulator for gate design purpose using the learning algorithms and to verify the gate operation, and investigated the robustness of the gate under distortion and random noise. S. P. verified the gate operation using QPT and investigated the effect of decoherence on the gate. S.D. and S.P. wrote the manuscript. All authors commented on the manuscript. A. M. supervised the project.